*Letter to the Editor*
# Radial cluster lensing

## A simple differential equation describing arclet fields

T. Schramm and R. Kayser

Hamburger Sternwarte, Gojenbergsweg 112, D-21029 Hamburg-Bergedorf, Germany
e-mail: tschramm@hs.uni-hamburg.de

**Abstract.** Clusters of galaxies acting as gravitational lenses deform the images of background galaxies. For small galaxies the deformation is determined by the local properties of the lens (shear, convergence). Small circular shaped galaxies are seen as small ellipses. The orientation and axial ratio of the ellipses could then be used to trace the properties of the lens. We introduce and solve an ordinary homogeneous linear first order differential equation which describes completely the radial cluster lens. It can be used to reconstruct the lens mappeing from the measured axial ratios of lensed background galaxies.

## 1. Introduction

Several efforts have been made to extract information from the arclet fields of clusters of galaxies (Tyson et al. 1984, Tyson 1985, Kochanek 1990, Bartelmann & Weiss 1993). Typically a parametric model is fitted to give the observed orientations and axial ratios of a population of background galaxies. This population is normally not known so that certain statistical assumptions have to be made (Miralda-Escudé 1991). Recently a nonparametric approach was introduced where the density distribution was approximated by a Poisson-like equation, which was solved by Fourier transform (Kaiser & Squires 1992, 1994 ). The underlying assumption is, however, the same, namely that the arclets can be considered statistically as being images of *circular* sources.

An interesting special case is the cluster lens with pure radial symmetry. It is treated with an iterative, parametric technique by Breimer (1994). We show here that an explicit integral exists which reconstructs the lens mapping up to a constant which corresponds to a sheet of constant density.

## 2. The radial lens

A radial lens mapping from the lens plane $(x,y)$ to the source plane $(u,v)$ can be written

$$u = x - \frac{M(r)}{r^2}x = \left(r - \frac{M(r)}{r}\right)\frac{x}{r} = R(r)\frac{x}{r} \quad , \tag{1}$$

$$v = y - \frac{M(r)}{r^2}y = \left(r - \frac{M(r)}{r}\right)\frac{y}{r} = R(r)\frac{y}{r} \quad , \tag{2}$$

where

$$R(r) = \sqrt{u^2 + v^2} = r - \frac{M(r)}{r} \tag{3}$$

and

$$r = \sqrt{x^2 + y^2}. \tag{4}$$

$M(r)$ is the mass inside a circle with radius $r$ and the usual normalizations are used (see Schneider et al. 1992, p 230 ff, see also Appendix).

The local lensing is given by the jacobian of the lens mapping

$$J = \begin{pmatrix} u_x & u_y \\ v_x & v_y \end{pmatrix} \quad , \tag{5}$$

where the subscripts denote differentiation. For the radial lens mapping it reads

$$J = \begin{pmatrix} \frac{dR}{dr}\frac{x^2}{r^2} + \frac{R}{r}\frac{y^2}{r^2} & \frac{dR}{dr}\frac{xy}{r^2} - \frac{R}{r}\frac{xy}{r^2} \\ \frac{dR}{dr}\frac{xy}{r^2} - \frac{R}{r}\frac{xy}{r^2} & \frac{dR}{dr}\frac{y^2}{r^2} + \frac{R}{r}\frac{x^2}{r^2} \end{pmatrix} \quad . \tag{6}$$

to an ellipse with axial ratio given by the ratio of the eigenvalues of the Jacobian Eq. (6)

$$\lambda^2 - (u_x + v_y)\lambda + u_x v_y - u_y v_x = 0 \quad , \tag{7}$$
$$\lambda^2 - \lambda \mathrm{Tr} J + \det J = 0 \quad . \tag{8}$$

From Eq. (6) we find

$$\mathrm{Tr} J = \frac{\mathrm{d}R}{\mathrm{d}r} + \frac{R}{r} \quad , \tag{9}$$
$$\det J = \frac{\mathrm{d}R}{\mathrm{d}r}\frac{R}{r} \tag{10}$$

and for the eigenvalues

$$\lambda_1 = \frac{R}{r} \quad , \tag{11}$$
$$\lambda_2 = \frac{\mathrm{d}R}{\mathrm{d}r} \quad . \tag{12}$$

A small circle is therefore stretched *tangentially* by a factor $\lambda_1$ and *radially* by $\lambda_2$. Thus, the axial ratio $\varepsilon$ of the elliptical image of a small circle mapped by the local lens is given by

$$\varepsilon = \frac{\lambda_1}{\lambda_2} = \frac{R}{r}\left(\frac{\mathrm{d}R}{\mathrm{d}r}\right)^{-1} \quad . \tag{13}$$

Note that the tangential and radial critical curve is given by $\lambda_1 = 0$ and $\lambda_2 = 0$ (see e.g. Schneider et al. 1992, p 235) which leads formally to $\varepsilon \to 0$ and $\varepsilon \to \pm\infty$, respectively. In any case $\varepsilon$ changes its sign (up to higher order singularities) at these locations. Under these circumstances the $\epsilon$-function is completely determined by measurement of the axial ratios of the arclets and boundary conditions as $\varepsilon \to 1$ if $r \to \infty$.

With these considerations Eq.(13) can be interpreted as ordinary homogeneous linear first order differential equation

$$\frac{\mathrm{d}R(r)}{\mathrm{d}r} - \frac{1}{\varepsilon(r)}\frac{R(r)}{r} = 0 \quad , \tag{14}$$

which is solved by

$$R(r) = C\exp\left(\int \frac{1}{r\varepsilon(r)}\mathrm{d}r\right) \quad . \tag{15}$$

The appearance of the constant $C$ beautifully recovers the *magnification transformation* as introduced by Gorenstein et al. (1988) (see also Kayser 1990).

The mass distribution is then given by (see Eq. (3))

$$M(r) = r^2 - Rr \quad . \tag{16}$$

We introduced and solved a differential equation for the radial lens mapping. It could be used for two purposes: (a) for a better theoretical understanding of the lensing process, since it shows what *could* be determined from a cluster lens with radial symmetry *if* all images are images of small circular sources (complete description), and (b) it could be used for real deconvolution of the lens mapping if the radial field of axial ratios of the arclets are measured and if some statistical assumptions are made. However, our approach supports the idea that cluster lensing does *not* immediately trace the density distribution or the potential of the cluster but the lens mapping from which the latter can be derived.

## A. Appendix

The coordinates $(u,v)$ and $(x,y)$ in the lens equation Eqs. (1,2) are angular coordinates. The mass $M(r)$ is then defined as

$$M(r) = 2\int_0^r \sigma(r')r'\mathrm{d}r' \tag{A1}$$

where $\sigma(r)$ is the surface density in units of the critical density

$$\sigma_{\mathrm{crit}} = \frac{c^2}{4\pi G}\frac{D_{\mathrm{s}}}{D_{\mathrm{ds}}D_{\mathrm{d}}} \tag{A2}$$

and the $D_{i,j}$ are angular size distances.

*Acknowledgements.* We thank A. Dent, P. Helbig, S. Refsdal and U. Borgeest for interesting discussions and helpful comments. This work was in part supported by the *Deutsche Forschungsgemeinschaft*, DFG under Schr. 417/1.